# Sunspots are in many ways similar to terrestrial vortices

Georgios H. Vatistas
Department of Mechanical and Industrial Engineering
Concordia University
1455 DeMaissoneuve Blvd. West
Montreal Canada
H3G 1M8

## Abstract

In this letter we identify similarities amongst sunspots and terrestrial vortices. The dark appearance of the central part of any sunspot is currently justified by an anticipated cooling effect experienced by the ionized gas. However, it cannot single-handedly reconcile the halo that surrounds the penumbra, the subsequent second dim ring that could be possibly followed by a second halo. In antithesis, light refraction due to density variations in a compressible whirl can give reason for all of these manifestations. Certain data of Wilson's depression fit better the geometric depth profile of a two-celled vortex. The last provides a hurricane equivalent manifestation for the normal and reverse Evershed effect. There is compelling evidence that alike to atmospheric vortices sunspots do also spawn meso-cyclones.

1. Introduction

Sunspot activity is a temporal event (almost periodic) having an approximately eleven-year period. It is associated with strong ultraviolet radiation known to influence several terrestrial climatic parameters including the global average temperature (see for example Lassen & Friis-Christensen (1995)). The current cycle (number 24) started in January 2009, it is presently in the upswing phase, expected to crest in the summer of 2014.

The world is filled with vortices and the sun is not exempted. Broadening previous work on turbulent vortices (Vatistas (2006[a])) a similarity within the similarity shared by intense eddies was uncovered. The new information provides the means to compress the tangential velocity profiles of a diverse collection of vortices into one and thus indentify those that belong to the same family. It was not unexpected to find several technological and atmospheric vortices to be members of this group. But curiously enough, some relatively small-scale solar circulations with dimensions of about three granulation cells reported by Brandt *et al* (1988) and Simon & Weiss (1997) do obey the relationship. The most salient results are presented and briefly discussed in Fig. 1. Another solar phenomenon also linked to whirlpool-like action is the solar spot. Astronomers have noticed the cyclone-like structure of sunspots since the Victorian era. In this letter we identify commonalities between these and ordinary vortices. Our primary motivation to conduct the study was Hale's (1915) remark that sunspots are in many respects analogous to tornadoes. The last undertaking was further fortified by the previous confirmation that the rotation curves of some relatively smaller scale solar whirls are akin to terrestrial vortices.

The central part of a spot is usually disk-like and dark (known as the umbra) surrounded by a lighter ring-type shadow (penumbra). Because the dark sector of large sunspots is visible with unaided vision, humans must have noticed them since the dawns of civilization. The first written record of solar black marks is due to Chinese astronomers (about the 9[th] century B.C) followed by the first explicitly documented case by the ancient Greek philosopher Theophrastus (around the 4$^{th}$ century B.C). However, truly epistemeiological approaches to describe the phenomenon appeared after the advent of the telescope, in the early part of the 17$^{th}$ century. Galileo, in addition to the discovery of Jupiter's satellites, he also reported the presence of dark spots in the solar disk. From the times of Galileo to early nineteen hundreds several important accounts of the event ensued without however any convincing explanation of its causality. In this precarious, fragmented, and full of speculations approach, the cyclone-like structure of sunspots emerged as one of their established features (Annon (1883) and (1884)). Hale (1908) conducted the first studies of the sunspots in a rigorous scientific manner. His spectrogram examinations revealed the then newly discovered Zeeman (1897) effect to be present thus concluding that the spots were under the influence of strong magnetic fields. Subsequently, based on this and previous observations where back-spirals were clearly visible with also signs of counterclockwise and clockwise rotations of sunspots in the northern and southern hemispheres respectively, remarked that these: "… are vortex phenomena, analogous in many respects to tornado …"(Hale (1915)). The presence of the Evershed radial flow rendered Hale's Zeeman effect based justification for the swirling motion in the ionized gas uncertain, Thomas & Weiss (2008). A detailed account on the historical evolution in this topic can be found in the fine contributions of Thomas & Weiss (2008) and Abetti (1965).

2. Analysis

Sunspots are highly complex events involving in addition to intricate fluid mechanical aspects, the effects of electromagnetic forces, several unfamiliar physical properties, conditions, and processes. In view of the aforementioned close likeness of solar whirls, Brandt *et al* (1988), Hale (1915), to various terrestrial vortices compelled us to assume that the fluid and plasma fields evolve in a similar manner.

Past experimental work (Sterling *et al* (1987)) has shown that when uniform light is shined on the liquid vortex with an interface shown in Fig. 2 a[†], the shadows casted due to refraction of light on the image plane possess some singular characteristics. Close to the vortex center a dark disk appears, surrounded by a bright halo, Sterling *et al* (1987)), Fig. 2 b. Depending on conditions such as for example focusing length, a second dim ring encircled by a subsequent bright ring may also emerge (Berry & Hajanal (1983)), see Fig. 2 c. Beyond the first or the second bright ring, light intensity approaches the background value asymptotically with the radius. A theoretical study (Aboelkassem & Vatistas (2007)) revealed that there is a dim ring (Fig. 2 b) between the dark disk and the first bright ring, which has been also identified by our recent shadowgraph experiments shown in Fig. 2 c. In addition, analytical work suggests that between the two halos there should also be another dim ring evident in Fig. 2 c. Images of sunspots over a century of observations show a remarkable resemblance to the previous shadowgraphs. The northern hemisphere sunspot, in Hale's (1924) spectroheliograph (Fig. 2 e), the umbra is indeed surrounded by a bright ring. In the previous picture (or the sunspot in the southern hemisphere, not shown here) the dim penumbra was undistinguishable from the umbra, this region is however clear in the image of Rast *et al* (1999$^)$ obtained using the Precision Solar Photometric Telescope at Mauna Loa Solar Observatory, see their figure 1 b. There is also a double bright halo combination that appears hazily in Hale's (2006) spectroheliograph, shown in Fig. 2 f. Additionally, Ananthakrishnan (1953) reported the presence of an additional bright ring encircling the penumbra. Similar shadow outlines could also be produced if instead of illuminating a vortex interface one shines uniform light on a compressible vortex shown in Fig. 2 g where its density varies with the radius (see Vatistas (2006$^b$) and Vatistas & Aboelkassem (2006)). This resemblance is not surprising since it is long known that shallow water hydraulics and the two dimensional compressible gas flows are analogous (Landau & Lifshitz (1987)). The shadowgraph measurements of helicopter rotor tip compressible vortex of Bagai & Leishman (1993) attest to the similarity. In addition, Das & Ramanathan (1953) have reported on the observed radiation intensity variation across a large sunspot. Their curves are remarkably similar to the shadowgraph contrast measurements of helicopter rotor tip compressible vortex of Bagai and Leishman (1993).

---

[†] The previous experiments were conducted with light shined from the top. Current laboratory experiments revealed that the same results would be obtained if the free liquid surface were illuminated from below.

The prevailing theory as to why the central part of a sunspot is dark is attributed to the fact that this region is at a considerably lower temperature than the rest of the photosphere. Previous theoretical studies indicated that indeed the central part of a compressible vortex experiences a cooling effect (Vatistas (2006[b])), see Fig. 2 h. The lower temperature hypothesis could indeed rationalize the dark portion of the spot, but alone cannot support the emergence of the bright halo(s) without an extra supposition about the nature of things. In contrast, refraction caused by an actual depression or the analogous density variation can provide a plausible cause for most of the observed optical manifestations. The last hypothesis must also include other than light electromagnetic radiation and possibly sound.

In 1769 astronomer Alexander Wilson noticed that when sunspots were positioned at the edges of the sun appeared as saucer-like depressions. One-cell fluid vortex with an interface is known to develop an inverted bell-like depression in the free surface. The fact however that these dimples could be shaped as a saucer suggests that such vortices must be of the two-cell kind. Evidence supporting the last proposition is found in the observations presented in Solanki (2003) shown in Fig. 3 where the two-cell vortex type is seen to approximate better the geometric depth. Several geophysical vortices are of the one cell kind where near the earth's surface radial wind converges towards the axis of rotation, it deflects upwards and closes the circuit by moving outwards at higher altitudes. Mature tropical cyclones and strong tornadoes do also develop an eye characterized by a meridional flow pattern that consists of two-cells. The flow in the outer cell is similar to the previous type. However, within the inner cell there is a toroidal recirculation zone consisting from a downdraft at the vortex center, and an outflow at the lower elevations. The previous stream joins the inflow of the outer cell and then both move together upwards. At higher altitudes and in order to complete the circulation, the inner cell stream moves horizontally towards the center of rotation while in the outer cell the stream flows outwards. This phenomenon is indeed analogous to both the normal and the reverse Evershed effects observed in some sunspots.

There is yet another common trait between sunspots and terrestrial vortices. Observations made using land-based and airborne radars of several hurricanes revealed that in addition to a circular shape of the eye-wall of tropical cyclones waveforms due to meso-cyclones with wave numbers $N$ = 2, 3, 4, 5, and 6 ($N$ represents the number of meso-vortices) are also present (Lewis & Hawkins (1982), Schubert *et al* (1999)). The last is in agreement with laboratory studies on polygonal cores (Vatistas *et al* (2008)). Tornadoes do as well form orbiting secondary revolving eddies (Fujita *et al* (1972)) (suction vortices). It is also expected that waves of this nature should appear in the interior of sunspots. The triangular form in the observed sunspot images of Mathew *et al* (2004) is quite revealing (see their Fig. 1). In their Fig. 7 the main vortex is seen to give rise to Wilson's main depression. In addition, there are three additional smaller dimples superimposed into the base surface. These must be due to the local satellite vortices orbiting the central vortex. The refracted light on the combined Wilson surface produces the triangular shape ($N$ = 3) for the sunspot given by Mathew *et al* (2004) in their Fig. 1. An image of a two lobed sunspot can be found in the article by Rajaguru (2009) (Fig. 2), while higher number polygonal spot structures can be seen in the images of Spruit & Roberts (1983) and Morita *et al* (2010).

3. **Conclusions**

In this paper we have identified several similarities between common vortices and sunspots. Light refraction can account for sunspot manifestations such as the dark appearance of the umbra, the shadowy penumbra, the possible double bright halos, as well as the dark ring between them. Hurricane analogous events in one- and two-celled vortices could also provide a valuable insight for both the normal and reverse Evershed effects. There is also convincing evidence in past observations that the main sunspot vortex can spawn meso-vortices. The last event along with refraction could give reason to the polygonal appearance of some spots.

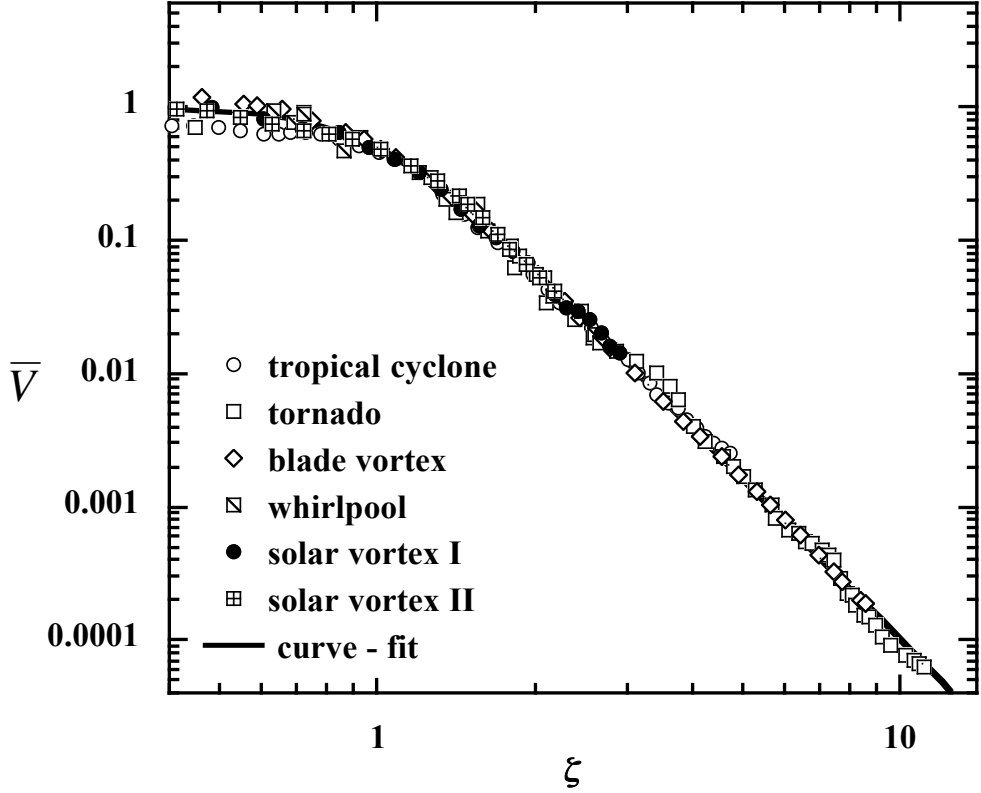

**FIGURE 1:** Curve-fitted observed tangential velocity radial distributions including a tropical cyclone Bonnie 1998 (Mallen *et al* (2005)) ($\beta$ = 2.565)), 30 May 1998 Spencer, South Dakota tornado (Wurman & Alexander (2005)) ($\beta$ = 2.086), a helicopter blade vortex (Bagai & Leishman (1993)) ($\beta$ = 1.370)), a whirlpool (Prichard (1970)) ($\beta$ = 1.000)), solar vortex I (Simon and Weiss (1997)) ($\beta$ = 2.050), and solar vortex II (Brandt *et al* (1988)) ($\beta$ = 1.000), along with equation (Vatistas (2006[a])) $\overline{V} = 1/(1+\zeta^4)$, where: $\overline{V} = (V/\xi)^{1/m}/(1+\beta)$, $\zeta = \beta^{1/4}\xi$, $V = \xi\{(1+\beta)/1+\beta\xi^4\}^m$, $m = (1+\beta)/4\beta$, $V = V_\theta/V_c$, $\xi = r/r_c$, $V_\theta$ is the tangential velocity, $V_c$ the tangential velocity at the core, $r$ is the radius, $r_c$ the core radius (or the radial distance from the center of rotation where the velocity attains its maximum value) and $\beta$ is a scaling dimensionless constant. The value of parameter $\beta$ is found in such a way as to minimize the square error of the theoretical values of $\overline{V}$ from the observed data. When $\beta$ = 1 the vortex is laminar-like, while $\beta$ > 1 represents turbulent vortices. It is interesting to note in passing that solar vortex I can be approximated by the laminar $n$ = 2 vortex (Vatistas *et al* (1991)), while II can only be curve-fitted by the turbulent equation.

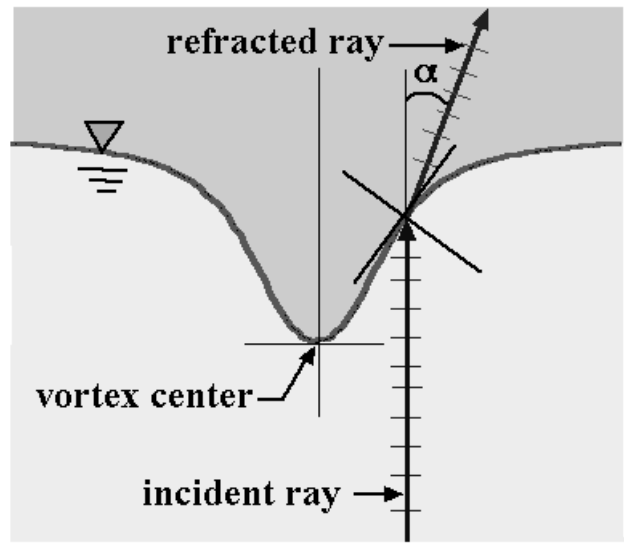

**(a)**

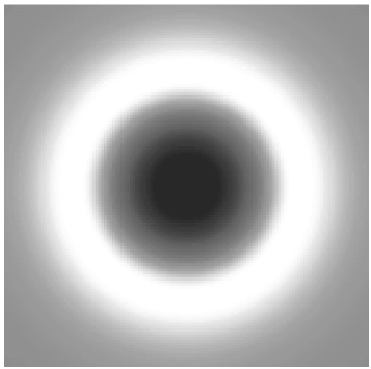

**(b)**

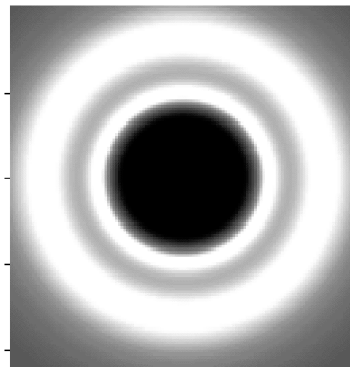

**(c)**

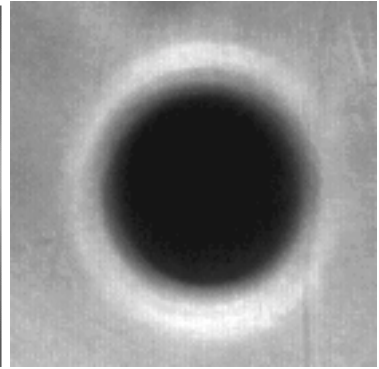

**(d)**

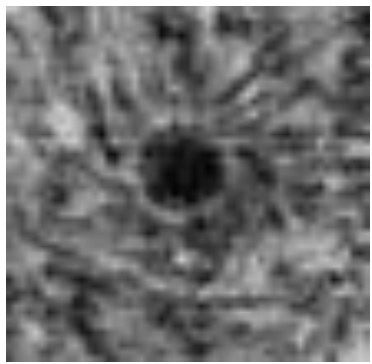

**(e)**

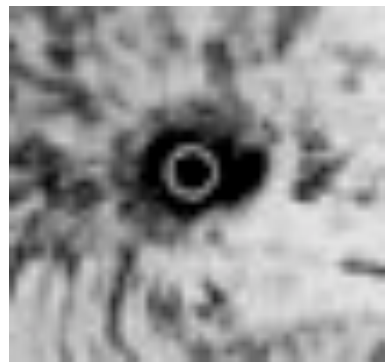

**(f)**

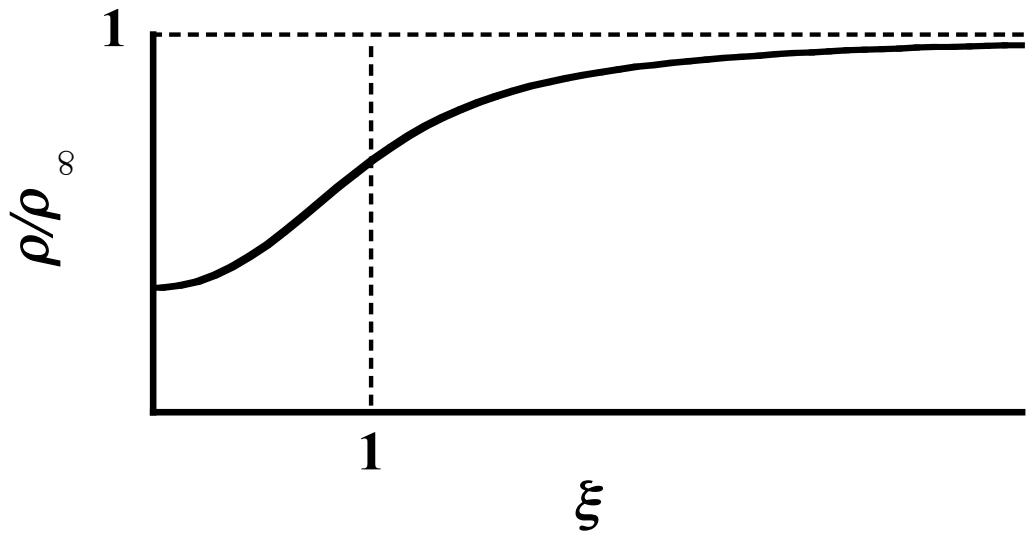

(g)

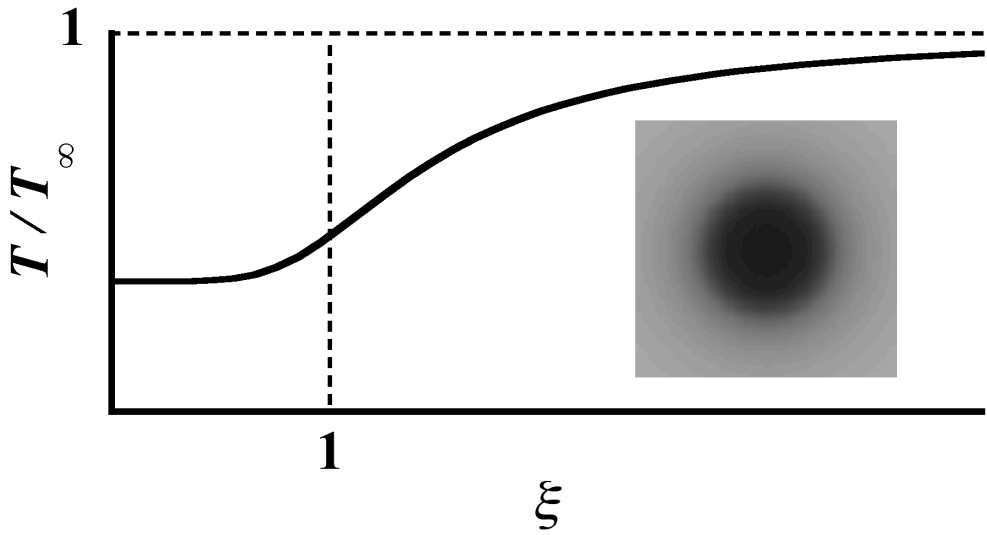

(h)

**FIGURE 2:** a Refraction of light in a fluid vortex with an interface. Shadowgraphs produced analytically b & c and the shadow of a water whirlpool photographed recently in our laboratory d. Northern hemisphere sunspot where the umbra appears to be surrounded by a bright ring is clearly evident in Hale's (1924) heliograph image e. In the previous picture the dim penumbra was undistinguishable from the dark umbra. The also possible double bright halo combination appears hazily in Hale's (2006) heliograph f. g Density variation in a compressible vortex (Vatistas (2006[b]), Vatistas& Aboelkassem (2006)). h Temperature profile in a compressible vortex (Vatistas (2006[b]), Vatistas & Aboelkassem (2006)), and the expected dark disk appearance due to the central cooling effect (assuming that luminosity is proportional to temperature). It is important to note that the pressure variation in a compressible vortex is similar to density and temperature.

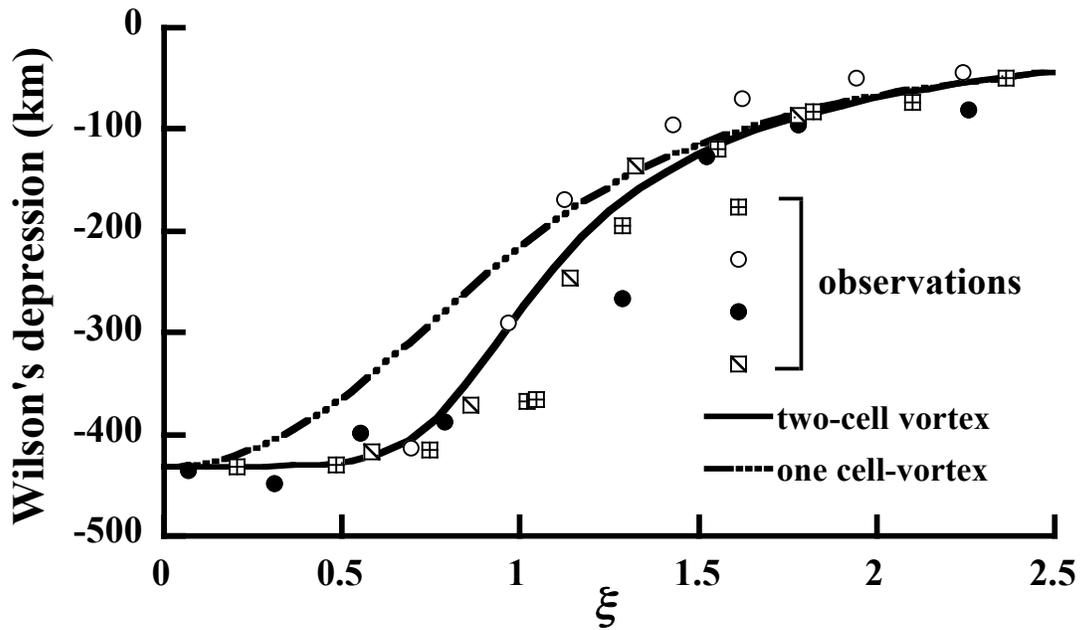

**FIGURE 2:** Observations by Solanki (2003) of Wilson's elevation profile along with curve fits for one- and two-celled vortices (Vatistas *et al* (1991), Vatistas (1998)). The two-cell vortex approximation (with curve fitting coefficients (Vatistas (1998))): $\kappa = 1.100$, $\beta = 0.600$, and $\eta = 0.625$) describes better the profile.